\documentclass[a4paper, times, 10pt,twocolumn]{article}
\usepackage[top=4.9cm,bottom=3.7cm,left=1.5cm,right=1.5cm]{geometry}
\usepackage{ICMLC}
\usepackage{times}
\usepackage{graphicx}
\usepackage{indentfirst}
\usepackage{latexsym}

\usepackage[autolinebreaks,useliterate]{mcode}
\usepackage{multirow}
\usepackage{hhline}
\usepackage{epstopdf}
\usepackage{subfigure}
\usepackage{amssymb}
\usepackage{amsmath}
\newcommand{\argmax}{\arg\!\max}

\pdfpagewidth=\paperwidth
\pdfpageheight=\paperheight
\begin{document}

\title{DOCUMENTS CLUSTERING BASED ON MAX-CORRENTROPY NONNEGATIVE MATRIX FACTORIZATION}  

\author{\bf{\normalsize{Le Li${^1}$, Jianjun Yang${^2}$, Yang Xu${^3}$, Zhen Qin${^3}$, Honggang Zhang${^3}$}}\\ 
\\
\normalsize{$^1$David R. Cheriton School of Computer Science, University of Waterloo, ON N2L3G1, Canada }\\
\normalsize{$^2$Department of Computer Science, University of North Georgia, Oakwood, GA 30566, USA }\\
\normalsize{$^3$Pattern Recognition and Intelligent System Laboratory, Beijing University of Posts and Telecommunications, Beijing, China}\\
\normalsize{E-MAIL: l248li@uwaterloo.ca, jianjun.yang@ung.edu, \{xj992adolphxy, qinzhenbupt\}@gmail.com, zhhg@bupt.edu.cn}\\
\\}


\maketitle \thispagestyle{empty}

\begin{abstract}
   {Nonnegative matrix factorization (NMF) has been successfully applied to many areas for classification and clustering. Commonly-used NMF algorithms mainly target on minimizing the $l_2$ distance or Kullback-Leibler (KL) divergence, which may not be suitable for nonlinear case. In this paper, we propose a new decomposition method by maximizing the correntropy between the original and the product of two low-rank matrices for document clustering. This method also allows us to learn the new basis vectors of the semantic feature space from the data. To our knowledge, we haven't seen any work has been done by maximizing correntropy in NMF to cluster high dimensional document data. Our experiment results show the supremacy of our proposed method over other variants of NMF algorithm on Reuters21578 and TDT2 databasets.}
\end{abstract}

\Section{Introduction}
A corpus is a collection of documents where each document is associated with a ground-truth topic that summaries the content of the document. Document clustering is the process that finds the correct label for the input document, such that this label should match with the ground-truth topic as much as possible. Such clustering makes it possible that automatically organizes millions of documents, websites, news, etc. into the multiple partitions, where documents within the same partitions share same topic. As a consequence, we can leverage this technique to different tasks, like document organization and browsing, corpus summarization, and document classification \cite{aggarwal2012survey}. 

Different types of algorithms have been used to cluster/classify the data (e.g. SVM \cite{zhou2010region} and pLSA \cite{zhou2013adaptive}). These algorithms have a variety of applications in different areas\cite{yu2010adaptive,sun2013novel,song2011image,xu2013cross,zhang2011distributed,sun2013space,sun2013mobile,shen2013virtual,shen2013layer,wang2012scimate}. Among those algorithms, we are especially interested in the nonnegative matrix factorization (NMF) method. NMF algorithm maps the original features into latent semantics space where each basis vector in the latent space represents a topic. More precisely, assuming each document is represented as a feature vector with $D$ dimension, and we have $N$ documents in the corpora, then we can form a $D*N$ matrix (denoted as $X$) to represent the whole corpora. NMF algorithm can decompose the $X$ into two low-rank nonnegative matrices, $H$ and $W$, such that the $X \approx HW$. One of the main benefits is its nature of dimension reduction without losing too much useful information. This decomposition has been shown its supremacy in many areas (e.g. bioinformatics \cite{wang2013non}). 

Much work has been done on applying NMF algorithms to document clustering \cite{xu2003document,shahnaz2006document}. However, most of them try to minimize the $l_2$ distance or KL divergence. Inspired by the recent work in \cite{wang2013non} that combines correntropy with NMF in cancer clustering, we propose a similar max-correntropy nonnegative matrix factorization algorithm (MCC) into the document clustering area. The work in \cite{wang2013non} is in line with ours in the way that both show the benefits of using this max-correntropy method for clustering. However, we are working on different areas. Meanwhile, the work in \cite{wang2013non} only examines the clustering performance on a limit number of topics (less than 10) and lower dimension data, while we systematically investigate its performance on more sophisticated clustering tasks with more documents, topics and higher dimension of data.

To achieve that, we implement the MCC algorithm and test its accuracy on the Reuters21578 and TDT2 corpora. We compare the MCC algorithm to classic loss functions ($l_2$ distance and KL divergence), as well as other variants of NMF algorithms. The results show that the proposed algorithm suppresses the rest methods on document clustering in both datasets. Moreover, we fully investigate how the MCC algorithm behaves when we keep increasing the number of topics (i.e. increasing the clustering difficulty). We find that although all algorithms' accuracies drop as we increase the number of topics, MCC algorithm is the most robust algorithm against the increment of the number of topics. 

The main contribution made in this paper is that, to our knowledge, this is the first work that factorizes the matrix from the perspectives of maximizing correntropy in document clustering. Besides that, we fully investigate its behaviors when faced with multiple topics' documents and high-dimension data. The results show the benefits of using the proposed MCC algorithm in document clustering.

\section{Related work}
\textit{K}-means is a classic clustering algorithm. This algorithm finds the closest cluster for each document by finding the smallest distance between the document and the existing clusters' centroids. The clusters' centroids are also updated at each iteration due to new cluster members. \textit{K}-means is based on the assumption that documents belonging to same topic should also be close to each other in the feature space. In a similar vein, Naive Bayes and Gaussian mixture model \cite{qin2012improving,baker1998distributional,liu2002document} are used based on different document distribution assumptions. One problem with these methods is that if the corpora properties don't following such assumptions, the performance of these algorithms may be at risk.

Latent Semantics Indexing (LSI) \cite{deerwester1990indexing} is the technique that converts the corpora from the original feature space into a latent semantics space. Each basis axis in the latent semantics space essentially represents one type of semantic information of the corpora. By doing so, each document is essentially a combination of multiple semantics information. Then we can apply the classic clustering algorithms on these new representations of documents in latent semantics space. One issue with this method is that the coefficients of the combination could be positive or negative. A negative coefficient is not such a natural way to interpret the document. Meanwhile, the bases that spanning the latent semantics space in some LSI algorithms, like Singular Vector Decomposition (SVD) \cite{furnas1988information}, are orthogonal, which means that every semantics bases are different from each other. However, in reality, this is not always the case.

Similar to LSI algorithms, NMF also maps the corpora into latent feature space. The differences are that: firstly, the bases in latent feature space don't need to be orthogonal. Also, each basis now corresponds to one topic of the corpora, which makes it very easy to determine the topic of document by simply choosing the largest component in the latent space. Meanwhile, every element in the two decomposed low-rank matrices are nonnegative. This additive combination makes it more natural to understand each document in an intuitive manner.

The benefits of using NMF in document clustering have been heavily investigated in many existed papers \cite{xu2003document,shahnaz2006document,lin2007projected,wang2012adaptive}. However, many of them are mainly targeting on minimizing the $l_2$ norm or KL divergence in the process of matrix decomposition. Correntropy-based decomposition methods have proved effective in many areas like cancer clustering \cite{wang2013non}, face recognition \cite{he2011maximum}, etc. Some other solutions can be found in \cite{sun2012unsupervised,cai2011graph}. However, we never find such technique in the document clustering research or be used for very high-dimension data with considerable number of clusters, which is another starting point of our work.

\section{Algorithm}
Assuming we have a matrix $X \in \mathbb{R}^{D \times N}$. NMF allows us to factorize $X$ into two nonnegative matrices $H \in \mathbb{R}^{D \times K}$ and $W \in \mathbb{R}^{K \times N}$, where the product $H*W$ approximates the original matrix X. Each column in $X$ is the feature vector of one document with $D$ elements. Thus, $X$ essentially represents the whole corpus with $N$ documents. Conventionally, we name $H$ as basis matrix that each column forms the basis vector of the semantic feature space, and $W$ as coefficient matrix. Hence, a document is further represented as the additive combination of weighted basis vectors in semantic space. $l_2$ norm and Kullback-Leibler (KL) divergence are two commonly-used measures of the similarity between original matrix $X$ and the product of $H$ and $W$. Based on different similarity measures, we are able to solve the factorization problem by minimizing the corresponding errors between $X$ and $H*W$. 

In this paper, we propose a new method to quantify the NMF by maximizing correntropy criteria in document clustering. Correntropy measures the generalized similarity between two random variables. More precisely, it models the expected differences between two random variables after mapping through kernel function. Without knowing the joint distribution of $X$ and $Y$, we can simply estimate the expectation by taking average (shown in  Equation~\ref{eq.correntropyEstimation}):
\begin{equation} \label{eq.correntropyEstimation}
\hat{V}_\sigma (x,y) = \frac{1}{D} \sum_{i=1}^D k_\sigma (x_i - y_i)
\end{equation}

where $k_\sigma (.)$ is the kernel function and $x_i$ and $y_i$ are the element in $X$ and $Y$, respectively.

Thus, instead of using $l_2$ distance or KL divergence, we try to find the basis matrix $H$ and coefficient matrix $W$, whose product $Y$ approximates $X$, by maximizing their correntropy on a feature-by-feature basis to allow for weighting each feature differently. For each feature, the kernel function can be calculated as:
\begin{equation}
k_\sigma \left(\sqrt{\sum_{n=1}^N (x_{dn} - \sum_{k=1}^K h_{dk} w_{kn})^2}\right)
\end{equation}

Hence, the correntropy maximization problem is expressed in Equation~\ref{eq.mccKernel}.
\begin{equation} \label{eq.mccKernel}
\max \limits_{h_{dk}>0,w_{kn}>0}\sum_{d=1}^D k_\sigma \left(\sqrt{\sum_{n=1}^N (x_{dn} - \sum_{k=1}^K h_{dk} w_{kn})^2}\right)
\end{equation}

To simplify the calculations without losing generality, we choose the Gaussian kernel function as $k_\sigma(.)$:

\begin{equation} \label{eq.gaussianKernel}
k_\sigma(x-y)=exp\left(-\gamma||x-y||^2\right)
\end{equation}

After substituting Equation~\ref{eq.gaussianKernel} back into Equation~\ref{eq.mccKernel}, the basis and coefficient matrices can be derived by solving:

\begin{equation}
\max \limits_{h_{dk}>0,w_{kn}>0} \sum_{d=1}^D exp \left(-\gamma \sum_{n=1}^N (x_{dn} - \sum_{k=1}^K h_{dk} w_{kn})^2\right)
\end{equation}

We introduce the convex conjugate function $\varphi(.)$ and auxiliary variables $\rho=[\rho_1,...,\rho_D]^\top$. Based on the theory of convex conjugate functions, the above optimization problem is equivalent to:

\begin{flalign}
&\max \limits_{H,W,\rho} \hat{F}(H,W,\rho) \nonumber \\
&s.t.\text{ } H \geq 0, W \geq 0&  \\
&\hat{F}(H,W,\rho)=\sum_{d=1}^D \left(\rho_d \sum_{n=1}^N (x_{dn}-\sum_{k=1}^K h_{dk}w_{kn})^2-\varphi(\rho_d) \right)& \nonumber
\end{flalign}

The optimization problem can be solved by Expectation-Maximization-like method. Starting from the initial value of $H$ and $W$, we compute $\rho$ in expectation step (E-step). Conditional on the $\rho$ value, we update the $H$ and $W$ values in maximization step (M-step). The process is called one iteration. This iterative process stops until it converges. The proposed MCC has a good convergence performance. We direct the readers to refer similar convergence proof in \cite{wang2013non}. We often assign $H$ and $W$ with random numbers to start the algorithm if we have no prior information about the distribution of data.

\textbf{E-step}: Starting from the estimated $H$ and $W$ from last M-step (or random values in the 1st iteration), $\rho$ of the \textit{t-th} iteration is computed as:

	\begin{flalign}
		&\rho_d^t=-g\left(\sqrt{\sum_{n=1}^N\left(x_{dn}-\sum_{k=1}^K h_{dk}^t w_{kn}^t\right)}, \sigma^t \right)&  \\
		&\text{where} \text{ } g(z,\sigma)=\sup \limits_{\varrho \in \mathbb{R^-}}\left(\varrho\frac{||z||^2}{\sigma^2}-\varphi(\varrho)\right)& \nonumber \\
		&\text{and} \text{ } \sigma^t = \sqrt{\frac{\theta}{2D} \sum_{d=1}^D \sum_{n=1}^N \left(x_{dn} - \sum_{k=1}^K h_{dk}^t w_{kn}^t \right)^2}& \nonumber
	\end{flalign}
	
\textbf{M-step}: Conditional on the new $\rho$ from last step, we compute the new basis and coefficient matrix, denoted as $H^{t+1}$ and $W^{t+1}$ respectively, by maximizing the object function:
	\begin{align}	
		&(H^{t+1},W^{t+1}) \nonumber \\
		= &\argmax_{H,W} \sum_{d=1}^D \left( \rho_d^t \sum_{n=1}^N \left(x_{dn}-\sum_{k=1}^K h_{dk} w_{kn} \right) ^2 \right) \nonumber \\
		= &\argmax_{H,W} Tr[(X-HW)^\top diag(\rho^t)(X-HW)] \nonumber \\
		= &Tr[X^\top diag(-\rho^t) X]-2Tr[X^\top diag(-\rho^t) HW] \nonumber \\
		&+Tr[W^\top H^\top diag(-\rho^t)HW] \nonumber \\
		&s.t. H \geq 0, W \geq 0& \nonumber
	\end{align}
where $Tr(.)$ means the trace of the matrix.
	
We apply the Lagrange method to solve the optimization problem. Let the elements of matrices $\Phi=[\phi_{dk}]$ and $\Psi=[\psi_{kn}]$ be the corresponding Lagrange multipliers for the nonnegative conditions of $h_{dk} \geq 0$ and $w_{kn} \geq 0$. Then we can express the Lagrange optimization problem as:
\begin{align}
L &= Trac[X^\top diag(-\rho^t) X]-2Trac[X^\top diag(-\rho^t) HW] \nonumber \\
&+Trac[W^\top H^\top diag(-\rho^t) HW] + Trac[\Phi H^\top] \nonumber \\
&+ Trac[\Psi H^\top]
\end{align}

The partial derivatives of $L$ with respect to $H$ and $W$ are:
\begin{align}
\frac{\partial L}{\partial H}&=-2diag(-\rho^t)XW^\top + 2diag(-\rho^t)HWW^\top+\Phi \nonumber \\
\frac{\partial L}{\partial W}&=-2H^\top diag(-\rho^t)X+2H^\top diag(-\rho^t)HW+\Psi \nonumber 
\end{align}

Set them to 0 based on Karush-Kuhn-Tucker optimal conditions, we have:
\begin{align}
&-(diag(-\rho^t)XW^\top)_{dk}h_{dk}+(diag(-\rho^t)HWW^\top)_{dk}h_{dk}=0\nonumber \\
&-(H^\top diag(-\rho^t)X)_{kn}w_{kn}+(H^\top diag(-\rho^t)HW)_{kn}h_{kn}=0 \nonumber
\end{align}

Hence, the basis matrix $H$ and coefficient matrix $W$ can updated as shown in Equation~\ref{eq:updateH} and Equation~\ref{eq:updateW}.
\begin{equation} \label{eq:updateH}
h_{dk}^{t+1} \leftarrow h_{dk}^t \frac{(diag(-\rho^t)XW^{t^\top})_{dk}}{(diag(-\rho^t)H^tW^tW^{t^\top}_{dk})_{dk}}
\end{equation}
\begin{equation} \label{eq:updateW}
w_{kn}^{t+1} \leftarrow w_{kn}^t \frac{(H^{{t+1}^\top}diag(-\rho^t)X)_{kn}}{(H^{{t+1}^\top}diag(-\rho^t)H^{t+1}W^t)_{kn}}
\end{equation}


\section{Experiment settings}
We test the MCC algorithm on two datasets: Reuters21578 \footnote{http://kdd.ics.uci.edu/databases/reuters21578/reuters21578.html} and TDT2 \footnote{http://www.itl.nist.gov/iad/mig/tests/tdt/1998/}. These two datasets have been widely used in many places \cite{xu2003document,shahnaz2006document} for document clustering. Reuters21578 test collection contains 21578 documents from 135 topics in total. We exclude those documents that belong to more than 1 topics in our experiment since we are trying to cluster each document to one single topic. Meanwhile, we also exclude those topics with less than 5 documents. As a consequence, we use 9545 documents for 51 topics in our experiment. TDT2 dataset contains around 11201 documents for 96 topics. We also apply similar pre-processing ways to it. The largest 30 topics with 9394 documents are used.

We use tf-idf method to extract the feature for each document. Stopwords and stemming are applied. The number of elements for each document are 16777 and 36771 for Reuters21578 and TDT2, respectively. 

After matrix decomposition, the matrix $W$ with dimension $K*N$ is essentially the new representation of the corpora in the way that each column is the feature vector of one document after dimension deduction. And the new dimension of the feature vector is $K$ now. To evaluate the decomposition performance, we directly apply \textit{K}-means clustering method to cluster $W$ into $K$ clusters. \textit{K}-means will assign each document with a label. We compare the label from \textit{K}-means to the original ground-truth label to calculate the clustering accuracy. The accuracy is defined in Equation~\ref{eq:accu}.
\begin{equation} \label{eq:accu}
Accuracy = \sum\limits_{i=1}^{N} \delta(kmeans\_label_i, topic_i)/N
\end{equation}
where $\delta(kmeans\_label_i, topic_i)$ is the delta function, which returns 1 if $kmeans\_label_i = topic_i$; otherwise 0. To find the correspondence between the topic from ground-truth data and the label by \textit{K}-means,  we use Kuhn-Munkres algorithm \cite{kuhn1955hungarian}.
\section{Results}
\newcommand{\tabincell}[2]{\begin{tabular}{@{}#1@{}}#2\end{tabular}}
\begin{table*}[ht]
\centering
\caption{Clustering accuracies on Reuters21578 and TDT2 datasets} \label{tab.results}
\begin{tabular}{|c||c|c|c|c|c||c|c|c|c|c|}
\hline
\multirow{2}{*}{\tabincell{c}{Number of \\ clusters}} & \multicolumn{5}{|c||}{Reuters21578} & \multicolumn{5}{|c|}{TDT2}\\
\hhline{~----------}
 & MCC & L2 & K-L & GS-CLS & PG & MCC & L2 & K-L & GS-CLS & PG \\ \hline
2 & 0.911 & 0.839 & 0.871 & 0.894 & 0.838  & 0.927 & 0.824 & 0.845 & 0.881 & 0.863 \\ \hline 
3 & 0.907 & 0.671 & 0.813 & 0.866 & 0.769 & 0.886 & 0.761 & 0.754 & 0.788 & 0.740 \\ \hline 
4 & 0.898 & 0.625 & 0.719 & 0.866 & 0.637 & 0.823 & 0.702 & 0.695 & 0.710 & 0.591 \\ \hline 
5 & 0.890 & 0.596 & 0.675 & 0.864 & 0.585 & 0.758 & 0.683 & 0.616 & 0.666 & 0.574 \\ \hline 
6 & 0.863 & 0.567 & 0.632 & 0.845 & 0.567 & 0.729 & 0.618 & 0.583 & 0.647 & 0.509 \\ \hline 
7 & 0.816 & 0.515 & 0.610 & 0.795 & 0.513  & 0.698 & 0.544 & 0.574 & 0.640 & 0.422 \\ \hline 
8 & 0.799 & 0.509 & 0.599 & 0.788 & 0.448 & 0.651 & 0.531 & 0.558 & 0.621 & 0.400\\ \hline 
9 & 0.770 & 0.494 & 0.490 & 0.770 & 0.409 & 0.647 & 0.533 & 0.556 & 0.573 & 0.395 \\ \hline 
10 & 0.736 & 0.403 & 0.457 & 0.719 & 0.415 & 0.631 & 0.468 & 0.518 & 0.555 & 0.382 \\ \hline 
\end{tabular}
\end{table*}

One important parameter we need to control is the number of cluster $K$. Intuitively, the value of $K$ controls the way to decompose the matrix. More importantly, it determines the number of topics that NMF algorithm can handle. That's to say, if $K = 2$, then the NMF is a two-topic clustering problem. If $K>2$, then it's a multi-topic clustering problem. To fully investigate the efficiency of the MCC, we use the following candidate numbers of clusters: \{2-10, 20, 30, 40, 51\} for Reuters dataset and \{2-10, 15, 20, 25, 30\} for TDT2 corpus. We randomly select $K$ topics, and use all documents from those $K$ topics as the current run's testing documents. We repeat this process 20 times to reduce the potential effect of random errors on our experiment results since  the performance of NMF algorithm is affected by the initial values of the iterative process. The average of 20 runs is used as the output of each algorithm. 

\begin{figure}[ht]
    \centering
    \includegraphics[width=0.8\columnwidth]{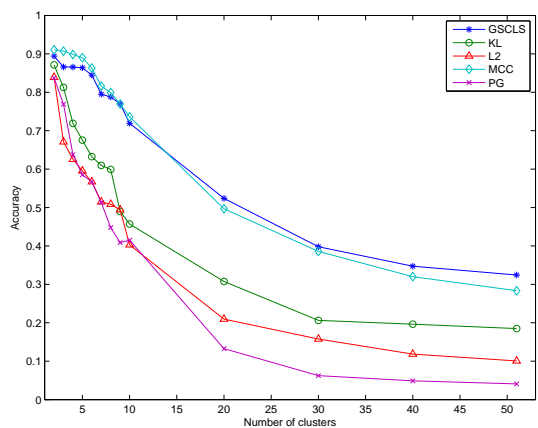}
    \caption{Accuracies on Reuters21578.}
    \label{fig:reuters}
\end{figure}

\begin{figure}[ht]
    \centering
    \includegraphics[width=0.8\columnwidth]{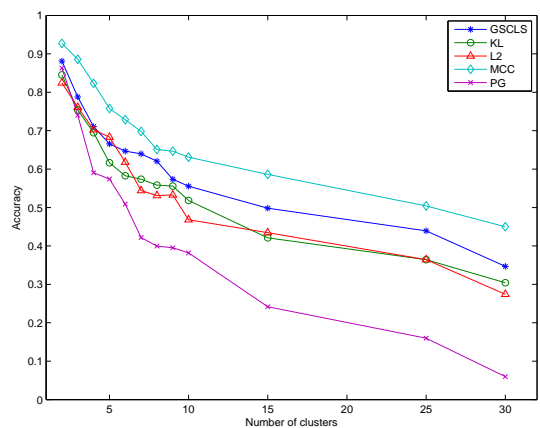}
    \caption{Accuracies on TDT2.}
    \label{fig:tdt2}
\end{figure}

Table~\ref{tab.results} summarizes the results of two datasets with less than or equal to 10 topics. Figure~\ref{fig:reuters} and Figure~\ref{fig:tdt2} demonstrate the accuracies on all chosen numbers of clusters. We firstly test the proposed MCC algorithm against two classic loss functions: $l_2$ distance and KL divergence. It's clear that MCC algorithm outperforms the $l_2$ distance and KL divergence in all cases of $K$ in two datasets. This shows the supremacy of the MCC algorithm against the others. One possible reason is that $l_2$ and KL distance are effective when dealing with linear separable data. However, if the data distribution is nonlinear manifold, it is considerably difficult for these two linear kernels to distinguish them.

Meanwhile, we observe that for all algorithms, the accuracy decreases as the number of clusters increases. Intuitively, more clusters inevitably increase the difficulties of finding the right label for each document. However, MCC is more robust to the increment of $K$, compared to other distance functions. 

We also compare MCC against two variants of NMF algorithms: gradient descent-constrained least squares (GSCLS) \cite{shahnaz2006document}, and Projected Gradient nonnegative matrix factorization (PG) \cite{lin2007projected}. Based on the results of two datasets, we can see that MCC suppresses the rest NMF algorithms when the number of clusters is smaller or equals to 10 on Reuters21578. When it comes to TDT2 dataset, MCC achieves the best performance in all cases, which shows the benefit of introducing the correntropy into the factorization process. One potential reason is that MCC can self-learn different kernels for different features. This adaptive learning property somehow further improves the performance of MCC when facing with nonlinear datasets (e.g. document collection). 

\section{Conclusion}
 \vspace{-4.1mm}
In this paper, we propose a new method to decompose the matrix into two low-rank matrices by maximizing the correntropy between them, such that we can easily and effectively use the decomposed matrix to cluster high-dimension data. We test the proposed MCC algorithm in the application of document clustering. We compare our proposed method to other loss functions and NMF algorithms. The results demonstrate the supremacy of our method on Reuters21578 and TDT2 corpora in terms of accuracy. In future, we will investigate the possibility of our proposed method in medical instrument\cite{anderson2013non}, mechanical instrument\cite{cui2012interpreting} and other related areas \cite{yang2014fairness,yang2013broadcasting,gao2012facial}.
\cite{zhou2013adaptive,li2014graph}

\section*{Acknowledgements}
This work was partially supported by National Natural Science Foundation of China under Grant No.61273217,  61175011 and 61171193, the 111 project under Grant No.B08004.

\nocite{ex1,ex2}
\bibliographystyle{latex8}
\bibliography{nmf-mcc-doc-icmlr}


\end{document}